\documentclass[twocolumn,showpacs,preprintnumbers,amsmath,amssymb]{revtex4}
\pdfoutput=1
\usepackage{graphicx}% Include figure files
\usepackage{dcolumn}% Align table columns on decimal point
\usepackage{bm}% bold math

\begin{document}

\newcommand{\BFA}{BaFe$_{2}$As$_{2}$}
\newcommand{\TCS}{ThCr$_{2}$Si$_{2}$}
\newcommand{\LFAOF}{LaFeAs(O$_{1-x}$F$_x$)}

%\preprint{APS/123-QED}

\title{Spin-density-wave anomaly at 140 K in the ternary iron arsenide BaFe$_{2}$As$_{2}$}

\author{Marianne Rotter, Marcus Tegel}
\author{Dirk Johrendt}
\email{johrendt@lmu.de} \affiliation{Department Chemie und Biochemie, Ludwig-Maximilians-Universit\"{a}t M\"{u}nchen, Butenandtstrasse 5-13 (Haus D), 81377 M\"{u}nchen, Germany}
\author{Inga Schellenberg, Wilfried Hermes, Rainer P\"{o}ttgen}
\affiliation{Institut f\"{u}r Anorganische und Analytische Chemie, Universit\"{a}t M\"{u}nster, Corrensstrasse 30,
D-48149 M\"{u}nster, Germany}

\date{\today}

\begin{abstract}

The ternary iron arsenide \BFA~with the tetragonal \TCS-type structure exhibits a spin density wave (SDW) anomaly at 140 K, very similar to LaFeAsO, the parent compound of the iron arsenide superconductors. \BFA~is a poor Pauli-paramagnetic metal and undergoes a structural and magnetic phase transition at 140 K, accompanied by strong anomalies in the specific heat, electrical resistance and magnetic susceptibility. In the course of this transition, the space group symmetry changes from tetragonal ($I4/mmm$) to orthorhombic ($Fmmm$). $^{57}$Fe M\"{o}ssbauer spectroscopy experiments show a single signal at room temperature and full hyperfine field splitting below the phase transition temperature (5.2 T at 77 K). Our results suggest that \BFA~can serve as a parent compound for oxygen-free iron arsenide superconductors.

\end{abstract}

\pacs{74.10.+v, 74.70.Dd, 71.27.+a, 75.30.Fv, 61.50.Ks, 61.05.cp, 33.45.+x}

\maketitle

%\section{\label{intro}Introduction}

\begin{figure}[h]
\includegraphics[width=75mm]{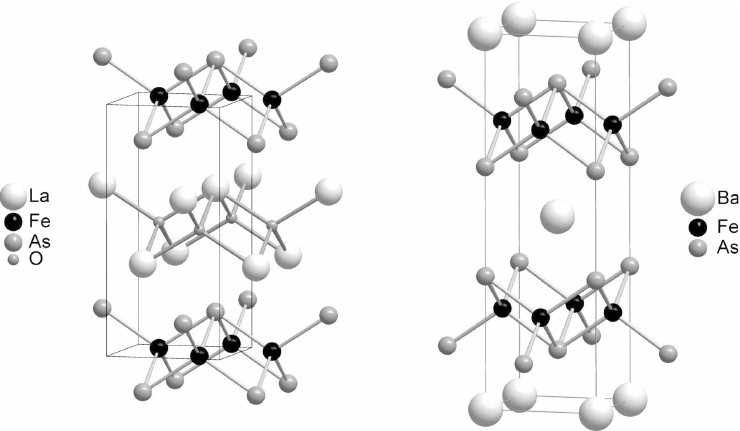}
\caption{\label{fig:Strukturvergleich}Crystal structures of LaFeAsO (left, ZrCuSiAs-type) and \BFA~(right, \TCS-type).}
\end{figure}

The recent discovery of superconductivity in doped iron-arsenide-oxides \cite{Hosono-2008} has heralded a new era in superconductivity research.\cite{Angew-2008} After the first report on \LFAOF~with a critical temperature ($T_C$) of 26 K, even higher transition temperatures up to 55 K in fluoride doped SmFeAs(O$_{1-x}$F$_x$) followed quickly. \cite{Sm-TC55} It is meanwhile accepted, that these materials represent the second class of high-$T_C$ superconductors after the discovery of the cuprates more than 20 years ago. \cite{Bednorz-1986} The parent compound LaFeAsO crystallizes in the tetragonal ZrCuSiAs-type structure (space group $P4/nmm$). \cite{Jeitschko-1974} Layers of edge-sharing OLa$_{4/4}$-tetrahedra alternate with layers of FeAs$_{4/4}$ tetrahedra along the $c$ axis, as shown on the left hand side in Figure~\ref{fig:Strukturvergleich}. This two-dimensional character of LaFeAsO involves different types of chemical bonding, which is strongly ionic in the LaO layers and rather covalent in the FeAs layers, respectively. From this purely ionic perception, we can assume a charge transfer according to (LaO)$^{\delta+}$(FeAs)$^{\delta-}$.

It is currently believed, that the superconductivity in doped LaFeAsO is, like in the cuprates, not of $s$-wave-type but intimately connected with magnetic fluctuations and a spin density wave (SDW) anomaly within the FeAs layers. \cite{Dong-SDW, Chen-SDW} Undoped LaFeAsO undergoes a SDW-driven structural phase transition round 150 K, associated with a reduction of the lattice symmetry from tetragonal to orthorhombic \cite{Nomura-2008} and anomalies in the specific heat, electrical resistance and the magnetic susceptibility. Antiferromagnetic ordering of the magnetic moments (0.36 $\mu_B$/Fe) was found below $T_N$ = 134 K by neutron scattering. \cite{Cruz-Neutrons} Electron doping with fluoride or oxygen deficiency, as well as hole doping with strontium suppresses the phase transition and the tetragonal phase becomes superconducting at $T_C$ = 25-41 K. \cite{Hosono-2008,LaFeAsOx-41K,LaSrFeAsx-25K} Hence, there is strong evidence that superconductivity in LaFeAsO emerges primarily from specific structural and electronic conditions of the (FeAs)$^{\delta-}$ layers.

Another well known structure type provides very similar conditions. The ternary iron arsenide \BFA~with the tetragonal \TCS-type structure (space group $I4/mmm$) \cite{Nagorsen-1980} contains practically identical layers of edge-sharing FeAs$_{4/4}$-tetrahedra, but they are separated by barium atoms instead of LaO sheets. Figure~\ref{fig:Strukturvergleich} shows both structures in comparison. In the very large family of \TCS-type compounds, superconductivity occurs scarcely and only at temperatures below 5 K. \cite{Shelton-1984} Examples are LaIr$_2$Ge$_2$, LaRu$_2$P$_2$, YIr$_{2-x}$Si$_{2+x}$ and BaNi$_2$P$_2$. \cite{Venturini-1985,Jeitschko-1987,Hagenmuller-1985,BaNi2P2-2008} In contrast to this, the related borocarbides are superconductors with higher $T_C$'s up to 26 K in YPd$_2$B$_2$C. \cite{Cava-1994} We suggested earlier, that superconductivity in \TCS-type compounds may be associated with direct interactions between the transition metal atoms, which can lead to structural instabilities. \cite{Johrendt-1997}

Regarding the previously described structural and electronic properties of LaFeAsO, we believe that among the \TCS-type compounds especially \BFA~is a very promising candidate for superconductivity for both structural and electronic reasons. In addition to the closely related geometry, also the electron counts of the FeAs layers in \BFA~and LaFeAsO are identical, because in both cases one electron is transferred to (FeAs) according to Ba$^{2+}_{0.5}$(FeAs)$^-$ and (LaO)$^+$(FeAs)$^-$, respectively. Apart from indeterminate magnetic data, \cite{Nagorsen-1983} no physical properties of \BFA~are known so far. In this paper, we report on a structural and magnetic phase transition at 140 K, specific heat, resistivity and magnetic measurements as well as $^{57}$Fe M\"{o}ssbauer spectroscopy of \BFA. We can show that the structural, electronic and magnetic properties of \BFA~and LaFeAsO are remarkably similar, which renders \BFA~a potential new parent compound for oxygen-free superconductors based on the \TCS-type structure.

%Synthese

\begin{figure}[h]
\includegraphics[width=75mm]{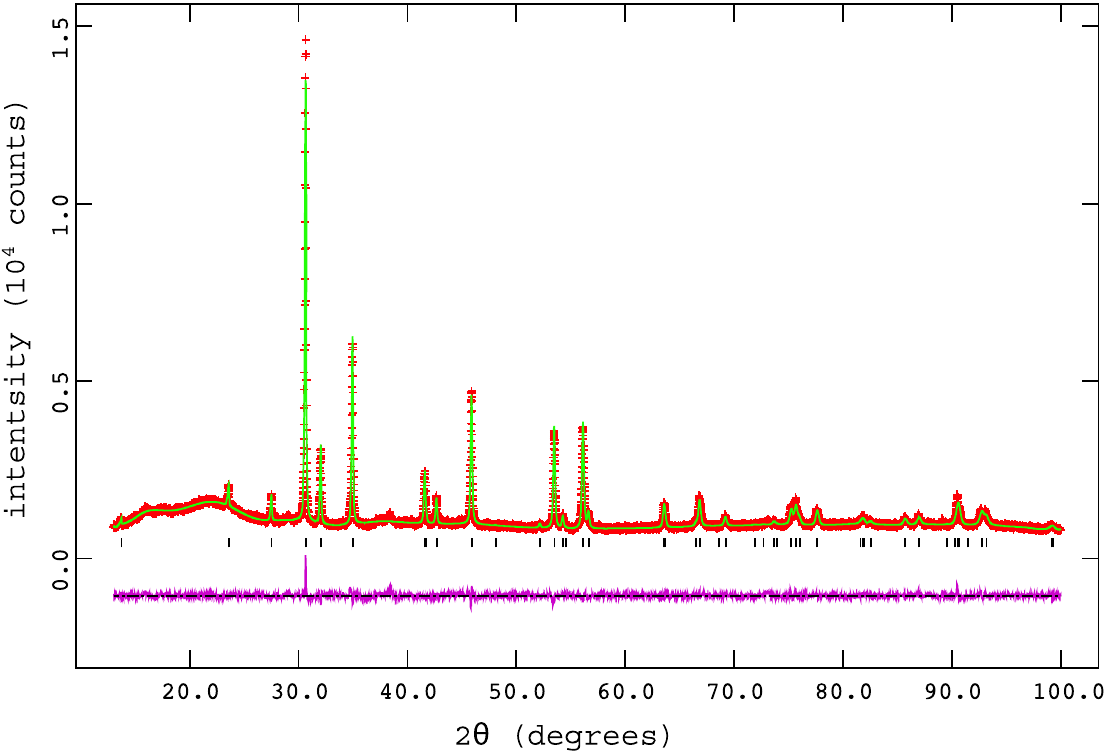}
\caption{\label{fig:Rietveld_297}(Color online) X-ray powder pattern (+) and Rietveld fit (-) of \BFA~at 297 K (space group $I4/mmm$).}
\end{figure}

\BFA~was synthesized by heating a mixture of distilled barium-metal, H$_2$-reduced iron-powder and sublimed arsenic at a ratio of 1.05:2:2 in an alumina crucible, which was sealed in a silica tube under an atmosphere of purified argon. The mixture was heated to 1123 K at a rate of 50 K/h, kept at this temperature for 10 h and cooled down to room temperature. The reaction product was homogenized in an agate mortar and annealed at 1173 K for 25 h. After cooling, the sample was homogenized again, pressed into a pellet ($\varnothing$=5mm, 1mm thick) and sintered at 973K. The obtained black crystalline powder of \BFA~is stable in air.
%
% X-ray at room temperature
%
Phase purity was checked by X-ray powder diffraction using a Huber G670 Guinier imaging plate diffractometer (Cu-$K_{\alpha_{1}}$ radiation, Ge-111 monochromator), equipped with a closed-cycle cryostat. Rietveld refinements of \BFA~were performed with the GSAS package \cite{GSAS} using Thompson-Cox-Hastings functions with asymmetry corrections as reflection profiles. \cite{Finger-Cox-Jephcoat} Figure~\ref{fig:Rietveld_297} shows the pattern of \BFA~, which could be completely fitted with a single phase. Impurities are, if at all, less than 1 \%.

\begin{figure}[h]
\includegraphics[width=65mm]{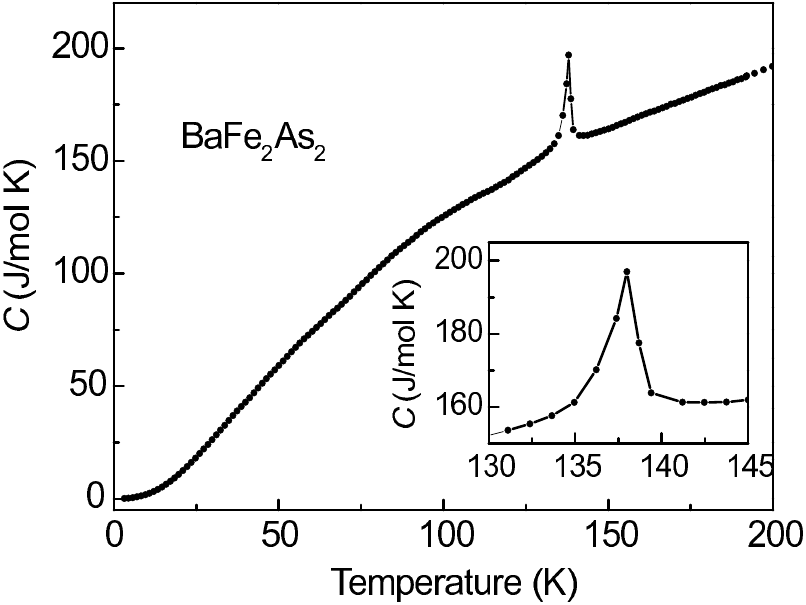}
\caption{\label{fig:SpecHeat}Specific heat of \BFA~vs. temperature.}
\end{figure}

%
% Spezifische Wärme
%
In order to check for a phase transition at about 150 K, as known from LaFeAsO, we measured the specific heat between 3 and 200~K by a relaxation-time method in a Physical Properties Measurement System (PPMS, Quantum Design Inc.). As it can be clearly seen from Figure~\ref{fig:SpecHeat}, we find a pronounced anomaly of $C_p(T)$ at $\approx$ 140 K. The characteristic $\lambda$-like shape of the peak points to a second order transition, as it is typical for magnetic ordering or a displacive structural change. From the inflection point of the $\lambda$-anomaly, we extracted the transition temperature of 139.9$\pm$0.5 K. In the low-temperature region the specific heat is of the form $C_p = \gamma T + \beta T^3$. The Debye temperature can be estimated from the equation $\beta = (12\pi^4nk_B)/(5\Theta_D^3)$, where $n$ is the number of atoms per formula unit. From a $C_p/T$ vs. $T^2$ plot between 3.1 and 14 K, we determined $\gamma$ = 16(2) mJ $K^{-2}$ mol$^{-1}$, $\beta$ = 2.0(5) mJ K$^{-4}$ mol$^{-1}$, and $\Theta_D$ =  134(1) K.

% Xray bei tiefen Temperaturen

\begin{figure}[h]
\includegraphics[width=75mm]{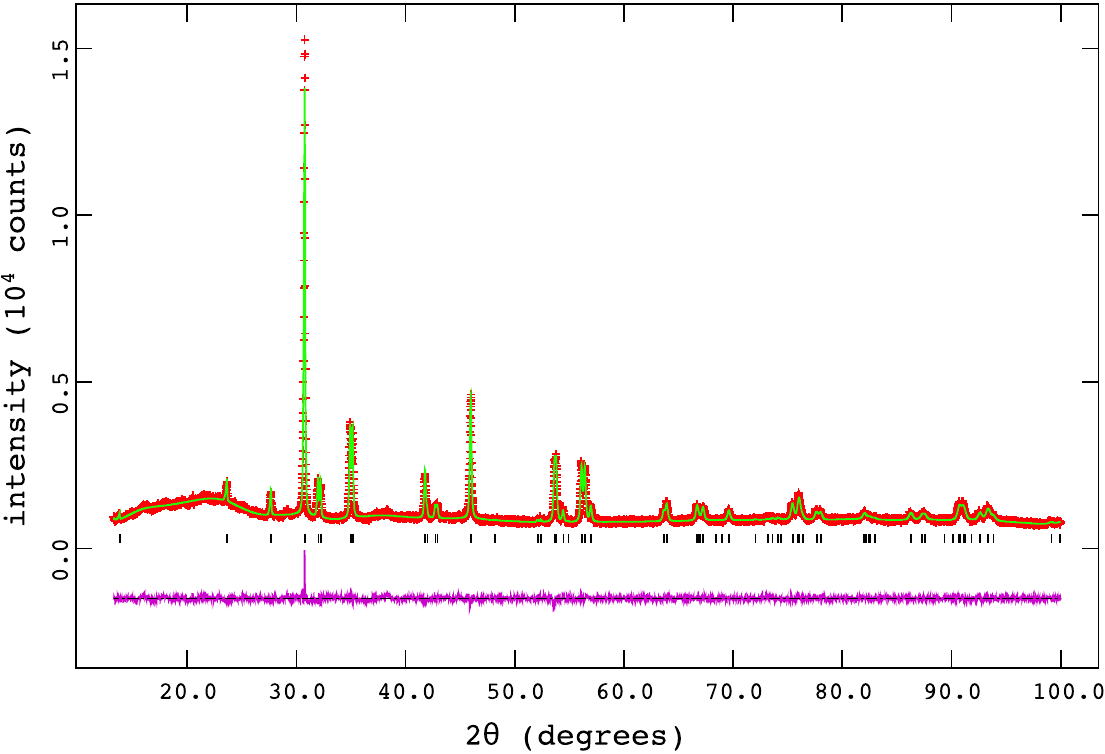}
\caption{\label{fig:Rietveld_20}(Color online) X-ray powder pattern (+) and Rietveld fit (-) of \BFA~at 20 K (space group $Fmmm$).}
\end{figure}

\begin{figure}[h]
\includegraphics[width=80mm]{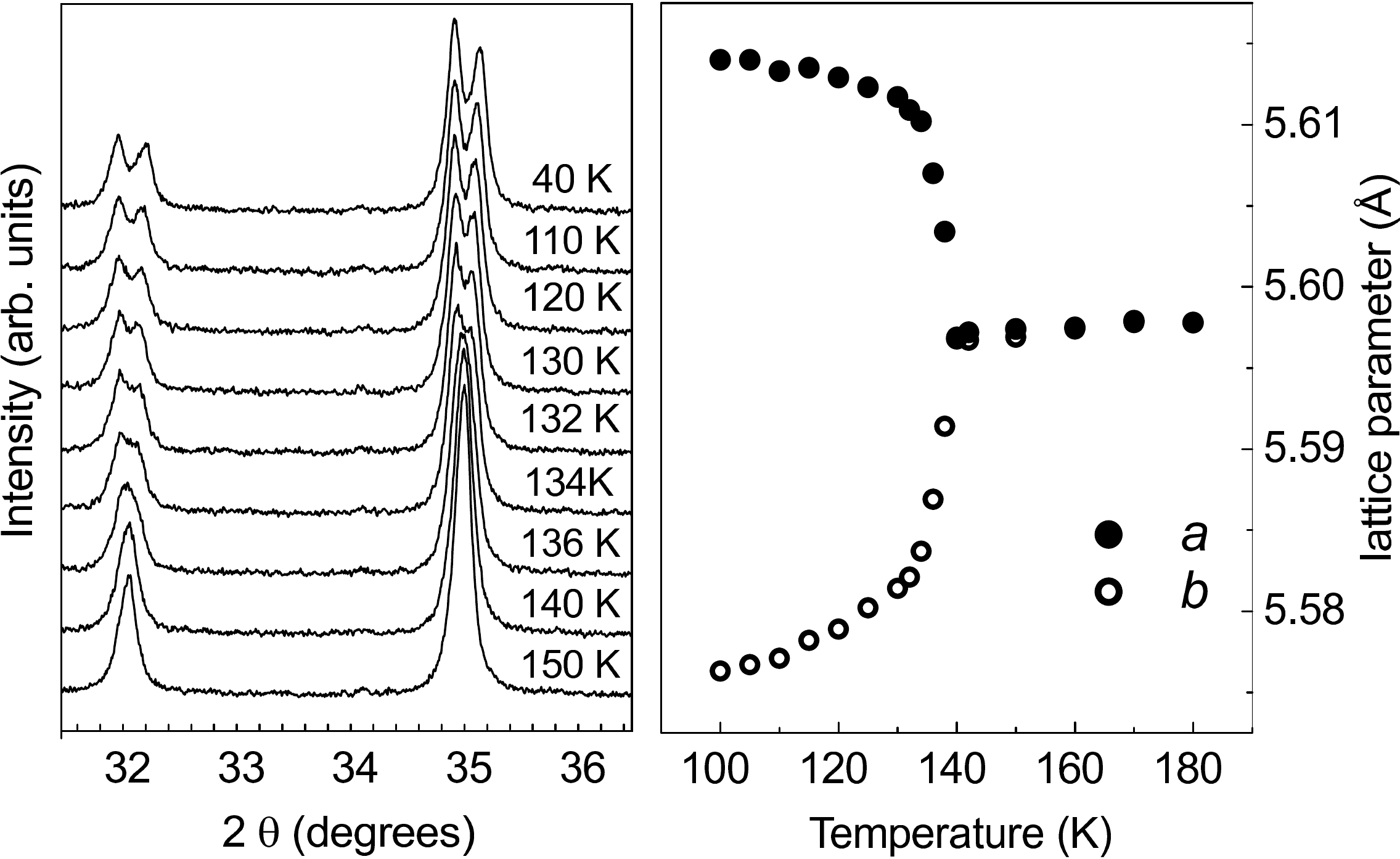}
\caption{\label{fig:Lattice}Splitting of the 110 and 112 reflections and variation of the lattice parameters with temperature. Values for the tetragonal phase above 140 K are multiplied by $\sqrt{2}$ for comparability.}
\end{figure}

Subsequently we have recorded X-ray powder patterns of \BFA~between 297 and 20 K. Several reflections broaden below 140 K and clearly split with further decreasing temperature. The patterns below 136 K were indexed with an orthorhombic $F$-centered unit cell ($a_{ortho} = \sqrt{2} \cdot a_{tetra}-\delta$; $b_{ortho} = \sqrt{2}\cdot b_{tetra}+\delta$; $c_{ortho} \approx c_{tetra}$; $\delta \approx$ 5 pm ). The low temperature data could be refined in the space group $Fmmm$. Figure~\ref{fig:Rietveld_20} shows the Rietveld fit of the data at 20 K. The continuous transition of the pattern between 150 and 40 K as well as the variation of the lattice parameters is depicted in Figure~\ref{fig:Lattice}. The space group $Fmmm$ is a subgroup of $I4/mmm$, thus a second order phase transition is in agreement with our data. In terms of group theory, this transition is \textit{translationengleich} with index 2 ($I4/mmm~\xrightarrow{t2}~Fmmm$). Crystallographic data are summarized in Table~\ref{tab:Crystallographic}. The main effect of the phase transition appears in the Fe--Fe distances, where four equal bonds of 280.2 pm length split into two pairs of 280.8 and 287.7 pm length. This supports the idea, that the Fe--Fe interactions are strongly correlated with the SDW anomaly and may play a certain role for the properties of \BFA.

\begin{table}[h]
\caption{\label{tab:Crystallographic} Crystallographic data of \BFA}
\begin{ruledtabular}
\begin{tabular}{lll}
 Temperature (K) & 297    & 20 \\
 Space group & $I4/mmm$           & $Fmmm$ \\
 \textit{a} (pm) & 396.25(1)        & 561.46(1) \\
 \textit{b} (pm) & $=a$             & 557.42(1) \\
 \textit{c} (pm) & 1301.68(3)       & 1294.53(3) \\
 \textit{V} (nm$^{3}$) & 0.20438(1) & 0.40514(2) \\
 \textit{Z} & 2                   & 4\\
 %$\rho_{calc}$ (g/cm$^3$) & 6.48  & 6.54\\
 data points & 8700               & 8675 \\
 reflections & 50                 & 74 \\
 atomic parameters & 4            & 4 \\
 profile parameters & 4           & 4 \\
 %background parameters & 36       & 36 \\
 %other parameters & 5             & 6 \\
 \textit {d} range & $0.979 - 6.508$ & $0.981 - 6.473$ \\
 R$_\text{P}$, \textit{w}R$_\text{P}$ & 0.0273, 0.0358 & 0.0283, 0.0365\\
 R$(F2)$, $\chi2$ & 0.0522, 1.431 & 0.0576, 1.392\\
 %GooF & 1.20 & 1.18 \\
Atomic parameters: \\
 Ba & 2$a$ (0,0,0)                         &  4$a$ (0,0,0)\\
    & $U_{iso} = 95(5)$                    & $U_{iso} = 69(5)$            \\
 Fe & 4$d$ ($\frac{1}{2},0,\frac{1}{4}$)   &  8$f$ ($\frac{1}{4},\frac{1}{4},\frac{1}{4}$)\\
    & $U_{iso} = 57(6)$                   & $U_{iso} = 64(4)$\\
 As & 4$e$ (0,0,$z$)                       &  8$i$ (0,0,$z$)  \\
    & $z$ = 0.3545(1)                    &  $z$ = 0.3538(1) \\
    & $U_{iso} = 99(5)$                  & $U_{iso} = 65(5)$\\
Bond lengths (pm):\\
Ba--As  &  338.2(1)$\times$8          & 336.9(1)$\times$4, 338.5(1)$\times$4 \\
Fe--As  &  240.3(1)$\times$4          & 239.2(1)$\times$4\\
Fe--Fe  &  280.2(1)$\times$4          & 280.7(1)$\times$2, 278.7(1)$\times$2\\
Bond angles (deg):\\
As--Fe--As &  111.1(1)$\times$2       & 111.6(1)$\times$2\\
           &  108.7(1)$\times$4       & 108.7(1)$\times$2, 108.1(1)$\times$2\\
\end{tabular}
\end{ruledtabular}
\end{table}

\begin{figure}[h]
\includegraphics[width=75mm]{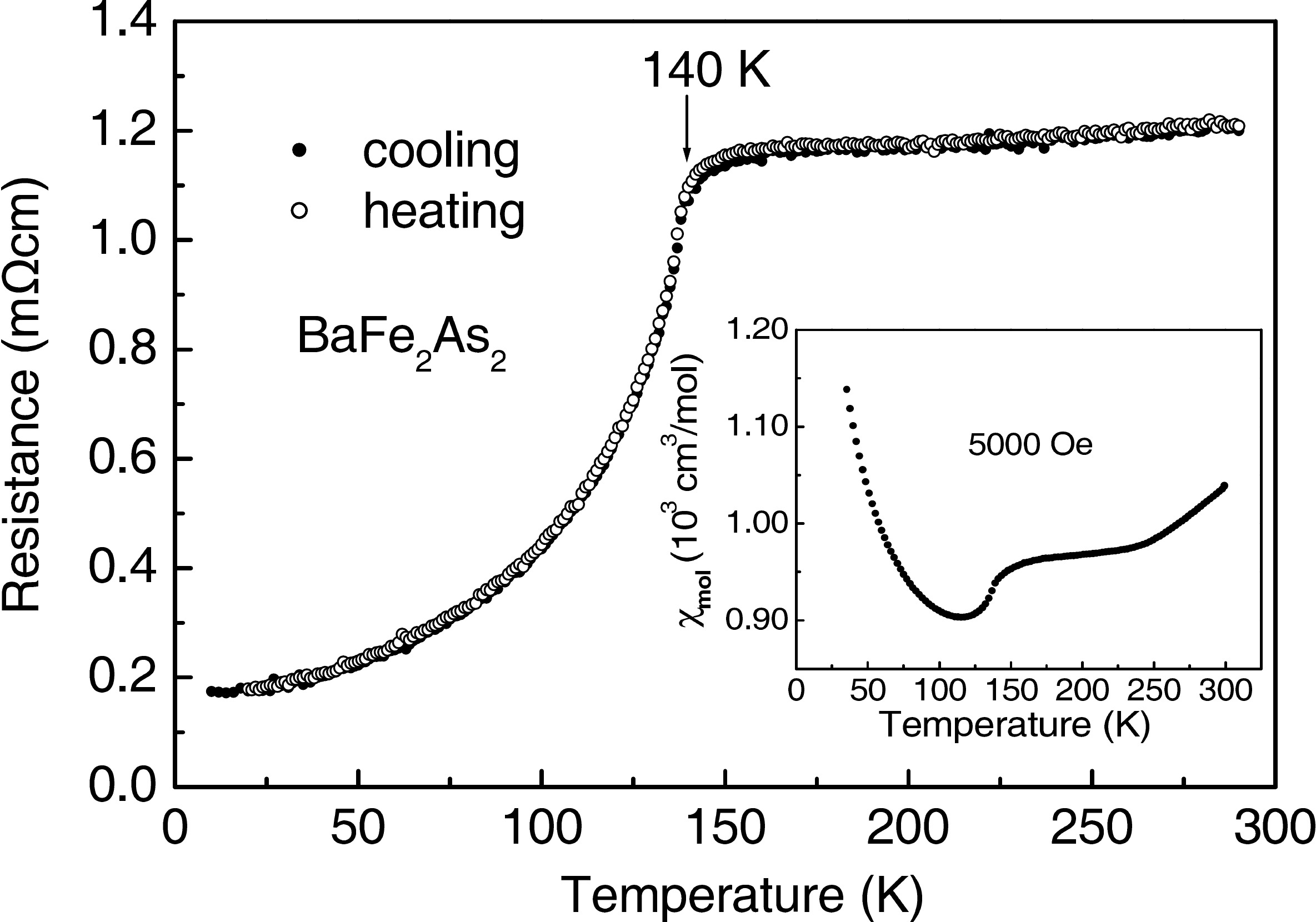}
\caption{\label{fig:rho+chi}$dc$ electrical resistance of \BFA~($I = 100 \mu$A). Insert: magnetic susceptibility measured at 0.5 T).}
\end{figure}

So far, our results clearly prove a structural distortion in \BFA. The nature of this effect is completely analogous to that in LaFeAsO, where a transition from the tetragonal space group $P4/nmm$ to the orthorhombic space group $Cmma$ occurs. \cite{Nomura-2008} As mentioned above, the transition of LaFeAsO is driven by a spin density (SDW) instability within the iron layers and therefore causes anomalies in the electrical resistivity and magnetic susceptibility. We have measured these properties of \BFA~and found again very similar behavior. The temperature dependency of the $dc$ electrical resistance is depicted in Figure~\ref{fig:rho+chi}. \BFA~ is a poor metal with a relatively high resistance around 1 m$\Omega$cm at room temperature, which decreases only slightly on cooling. At 140 K, the resistance drops abruptly at first but then decreases monotonically to 0.2 m$\Omega$cm at 10 K. This behavior corresponds to undoped LaFeAsO, where the resistance is of the same magnitude at room temperature and drops in a similar fashion. \cite{Dong-SDW}

Finally we have investigated the general magnetic properties and the specific behavior at the phase transition. The magnetic susceptibility was measured with a SQUID magnetometer (MPMS-XL5, Quantum Design Inc.) at 0.5 T. \BFA~shows a weak and only slightly temperature-dependent paramagnetism, as it is typical for a Pauli-paramagnetic metal. Below 140 K $\chi$ drops at first but it increases again below 100 K. The latter fact may be attributed to traces of ferromagnetic impurities, which are not detectable by the X-ray powder method.

\begin{figure}
\includegraphics[width=50mm]{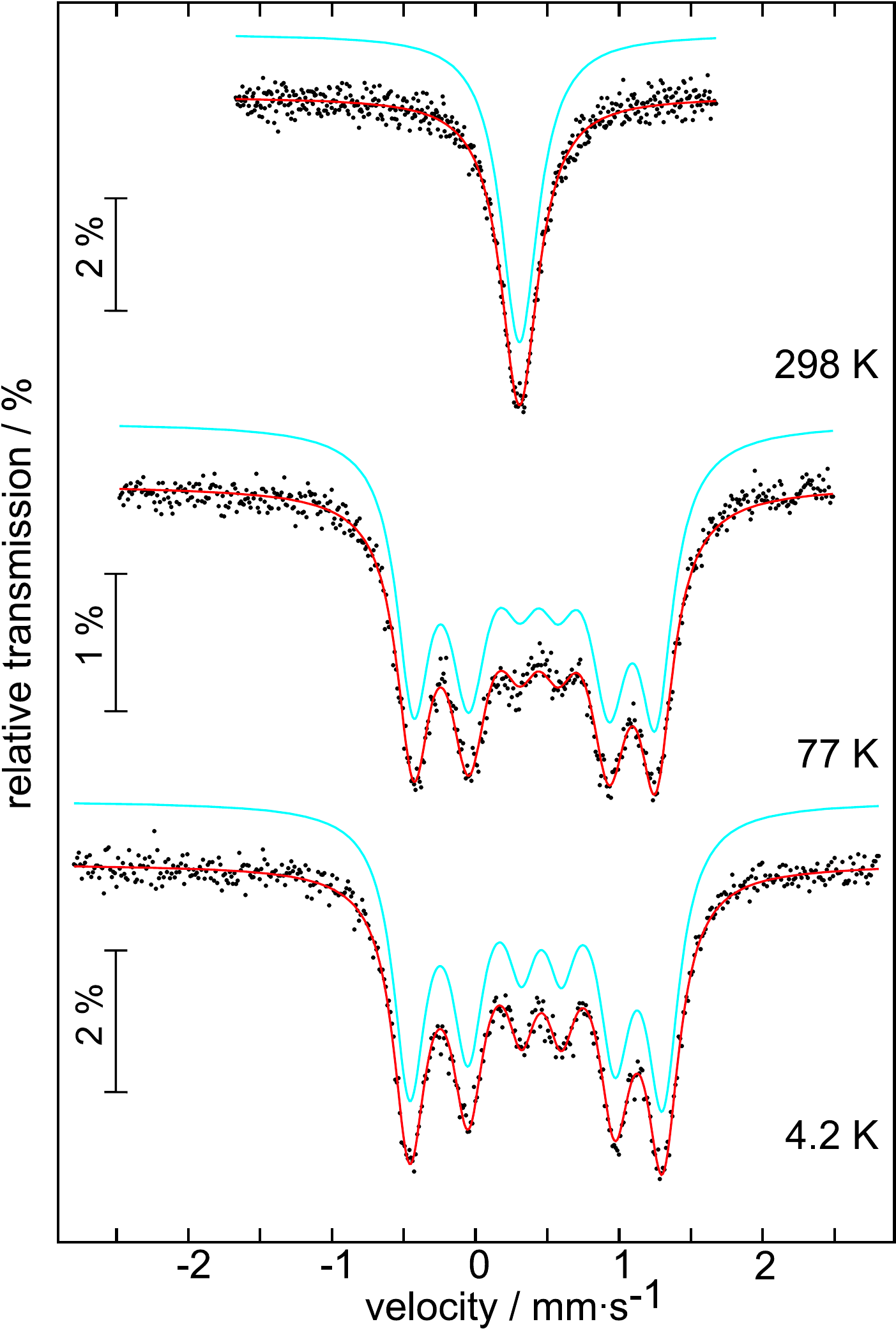}
\caption{\label{fig:MB}(Color online) $^{57}$Fe M\"{o}ssbauer spectra of \BFA~with transmission integral fits.}
\end{figure}

A $^{57}$Co/Rh source was available for $^{57}$Fe M\"{o}ssbauer spectroscopy investigations. The \BFA~sample was placed in a thin-walled PVC container at a thickness of about 4 mg Fe/cm$^2$. The measurements were performed in the usual transmission geometry at 298, 77, and 4.2 K, the source was kept at room temperature. $^{57}$Fe M\"{o}ssbauer spectra of \BFA~at 298, 77, and 4.2 K are shown in Figure~\ref{fig:MB} together with transmission integral fits. The corresponding fitting parameters are listed in Table~\ref{tab:MB-Data}. At room temperature we observed a single signal at an isomer shift of $\delta$ = 0.31(1) mm$\cdot$s$^{-1}$. Although the iron atoms have a non-cubic coordination by arsenic, there was no need to consider quadrupole splitting in the fits. The observed isomer shift is slightly smaller than in SmFeAsO$_{0.85}$ \cite{Felner-2008} and LaFeAsO. \cite{Kitao-2008,Klauss-2008} At 77 K, well below the transition temperature, we observe significant hyperfine field splitting with a hyperfine field value of 5.23(1) T, which is even slightly larger than the hyperfine field observed for LaFeAsO (4.86 T). \cite{Klauss-2008} A very small quadrupole splitting parameter of -0.03(1) mm/s was included in the fits. This parameter accounts for the small tetragonal-to-orthorhombic structural distortion. A similar value has been observed for the spin-density-wave system LaFeAsO below the transition temperature. \cite{Kitao-2008} The quadrupole splitting parameter slightly increases to -0.04(1) mm/s at 4.2 K (Table II). The hyperfine field at the iron nuclei is 5.47(1) T and the corresponding magnetic moment was estimated as 0.4$ \mu_B$ per iron atom.

\begin{table}[h]
\caption{\label{tab:MB-Data} Fitting parameters of $^{57}$Fe M\"{o}ssbauer spectroscopy measurements with \BFA.}
\begin{ruledtabular}
\begin{tabular}{lllll}
T (K) &  $\delta$ (mm$\cdot$s$^{-1}$) & $\Gamma$ (mm$\cdot$s$^{-1}$) & $\Delta E_Q$ (mm$\cdot$s$^{-1}$) &  $B_{hf}$ (T)\\
\hline
295 & 0.31(1)  &  0.32(1) &  --  &  -- \\
77  & 0.43(1)  &  0.33(2) &  -0.03(1) & 5.23(1)\\
4.2 & 0.44(1)  &  0.25(1) &  -0.04(1) & 5.47(1)\\
\end{tabular}
\end{ruledtabular}
\end{table}

In summary, we have shown that the properties of the ternary arsenide \BFA~with the \TCS-type structure are remarkably similar to LaFeAsO, the parent compound of the newly discovered superconductors. Both materials are poor metals at room temperature and undergo second order structural and magnetic phase transitions. The $^{57}$Fe M\"ossbauer data of \BFA~show hyperfine field splitting below 140 K, which also hints at antiferromagnetic ordering. Consequently, \BFA~exhibits the same SDW anomaly at 140 K as LaFeAsO at 150 K. Since this SDW instability is actually believed to be an important prerequisite for high-$T_C$ superconductivity in iron arsenides, our results strongly suggest that \BFA~can serve as a parent compound for another, oxygen-free class of iron arsenide superconductors with \TCS~type structure. All the signs are that superconductivity in \BFA~can be induced either by electron or hole doping. If this is the case, it would conclusively prove that superconductivity originates only from the iron arsenide layers, regardless of the separating sheets. On the other hand,  superconductivity in doped \BFA~would open new avenues to further high-$T_C$ materials in the large family of \TCS-type compounds.

\begin{acknowledgments}

We thank Dipl.-Chem. F. M. Schappacher for help with the M\"{o}ssbauer spectroscopy experiments. This work was financially supported by the DFG.

\end{acknowledgments}

\bibliographystyle{apsrev}

%\bibliography{Johrendt_PRBrc}

\end{document}